\documentclass[10pt, twocolumn, twoside]{IEEEtran}
\usepackage{amsmath, amssymb, amsfonts, amsthm}
\usepackage{graphicx, epstopdf, color, url}
\usepackage{bm, latexsym, multirow, booktabs}
\usepackage{algorithm, algpseudocode}
\usepackage{lscape}
\usepackage{grffile}
\usepackage{psfrag, subeqnarray}
\usepackage{cuted, stfloats}
\usepackage[acronym]{glossaries}
\usepackage{enumitem}
\usepackage{cite}
\usepackage{subcaption}
\usepackage{svg}
\usepackage{tabularx}

\begin{document}

\title{Single-Snapshot Gridless 2D-DoA Estimation for UCAs: A Joint Optimization Approach}

\author{
\IEEEauthorblockN{Salar Nouri\IEEEauthorrefmark{1}}\\
\IEEEauthorblockA{\IEEEauthorrefmark{1}School of Electrical and Computer Engineering, College of Engineering, University of Tehran, Tehran, Iran \\
Emails: \{salar.nouri\}@ut.ac.ir}
}

\maketitle
	
\begin{abstract}
This paper tackles the challenging problem of gridless two-dimensional (2D) direction-of-arrival (DOA) estimation for a uniform circular array (UCA) from a single snapshot of data. Conventional gridless methods often fail in this scenario due to prohibitive computational costs or a lack of robustness. We propose a novel framework that overcomes these limitations by jointly estimating a manifold transformation matrix and the source azimuth-elevation pairs within a single, unified optimization problem. This problem is solved efficiently using an inexact Augmented Lagrangian Method (iALM), which completely circumvents the need for semidefinite programming. By unifying the objectives of data fidelity and transformation robustness, our approach is uniquely suited for the demanding single-snapshot case. Simulation results confirm that the proposed iALM framework provides robust and high-resolution, gridless 2D-DOA estimates, establishing its efficacy for challenging array signal processing applications.
\end{abstract}

\begin{IEEEkeywords}
    Inexact Augmented Lagrangian (iALM), Gridless Estimation, Single-Snapshot DOA, 2D-DOA Estimation, Uniform Circular Array (UCA).
\end{IEEEkeywords}
\IEEEpeerreviewmaketitle
\newacronym{doa}{DOA}{Direction-of-Arrival}
\newacronym{radar}{RADAR}{Radio Detection and Ranging}
\newacronym{sonar}{SONAR}{Sound Navigation and Ranging}
\newacronym{music}{MUSIC}{MUltiple SIgnal Classification}
\newacronym{esprit}{ESPRIT}{Estimation of Signal Parameters via Rotational Invariance Techniques}
\newacronym{ssr}{SSR}{Sparse Signal Representation}
\newacronym{cs}{CS}{Compressive Sensing}
\newacronym{anm}{ANM}{Atomic Norm Minimization}
\newacronym{sdp}{SDP}{Semidefinite Program}
\newacronym{uca}{UCA}{Uniform Circular Array}
\newacronym{ula}{ULA}{Uniform Linear Array}
\newacronym{sip}{SIP}{Semi-Infinite Program}
\newacronym{ialm}{iALM}{Inexact Augmented Lagrangian Method}
\newacronym{iid}{i.i.d.}{independent and identically distributed}
\newacronym{snr}{SNR}{signal-to-noise ratio}
\newacronym{anm}{ANM}{Atomic Norm Minimization}
\newacronym{qcqp}{QCQP}{Quadratically Constrained Quadratic Program}
\newacronym{2d}{2D}{two-dimensional}
\newacronym{mlt}{MLT}{multilevel Toeplitz}
\newacronym{ura}{URA}{uniform rectangular array}
\newacronym{1d}{1D}{one-dimensional}
\newacronym{bcd}{BCD}{Block Coordinate Descent}
\newacronym{rmsae}{RMSAE}{Root Mean Square Angular Error}
\newacronym{lmi}{LMI}{Linear Matrix Inequality}
\newacronym{crb}{CRB}{Cramér-Rao Bound}
\section{Introduction}
\label{sec:intro}

\gls{doa} estimation is a cornerstone of array signal processing, with critical applications in \gls{radar}, \gls{sonar}, and wireless communications. While \gls{1d} \gls{doa} estimation is well-established, the joint estimation of azimuth and elevation angles in \gls{2d} \gls{doa} provides a far more complete spatial awareness. The \gls{uca} is a popular choice for \gls{2d}-\gls{doa} due to its full $360^\circ$ azimuth coverage and isotropic response \cite{schmidt1986multiple, roy2002esprit}. However, the problem becomes exceptionally challenging under the single-snapshot condition, where only a single vector of measurements is available \cite{van2002optimum}. This scenario is crucial for applications involving fast-moving targets or rapidly changing signal environments, yet it renders many classical estimation techniques ineffective.

Conventional high-resolution methods, such as \gls{2d}-\gls{music}, rely on the eigen-decomposition of the sample covariance matrix \cite{wang2016coarrays}. With a single snapshot, this matrix becomes rank-one, making it impossible to distinguish the signal and noise subspaces. While pre-processing techniques like spatial smoothing can create a higher-rank matrix, they do so at the cost of a reduced effective array aperture \cite{tang2013compressed}. To address these issues, methods based on \gls{cs} reformulate \gls{doa} estimation as a sparse recovery problem \cite{malioutov2005sparse}. However, grid-based techniques are fundamentally limited by basis mismatch, where their accuracy is compromised when true \glspl{doa} lie between predefined grid points. This motivated the development of gridless methods based on \gls{anm}, which operate on a continuous angular domain \cite{tang2013compressed}. For \gls{2d} problems, however, the canonical \gls{anm} formulation requires solving a large-scale \gls{sdp}, whose high computational complexity remains a significant barrier to practical application \cite{zheng2024ris, yang2016vandermonde, mathews2002eigenstructure}.

For \glspl{uca}, prized for their complete azimuthal coverage, a popular strategy is to map the complex \gls{uca} manifold to a simpler virtual array manifold via a linear transformation matrix, $\mathbf{T}$ \cite{mathews2002eigenstructure}. To solve the \gls{2d}-\gls{doa} problem, this requires forming a virtual \gls{ura} whose separable Kronecker structure is amenable to efficient estimators, such as ESPRIT. The success of this entire approach, however, hinges critically on the quality and robustness of the matrix $\mathbf{T}$.

Analytical methods, such as the well-known Davies transformation \cite{davies1965transformation}, are fundamentally ill-suited for this task. They are derived under ideal, noise-free conditions and are notoriously fragile; their construction involves inverting Bessel functions that are often near-zero, leading to severe noise amplification and unreliable estimates \cite{belloni2006beamspace}.

Recognizing this fragility, some research has focused on designing a better $\mathbf{T}$ as a separate, offline optimization problem \cite{labbaf2023robust}. These methods typically seek to minimize the Frobenius norm of $\mathbf{T}$ (to control noise gain) subject to manifold approximation error constraints. While an improvement, this offline approach is fundamentally limited. The resulting matrix is static and data-independent, unable to adapt to the specific signal and noise characteristics of the received snapshot. This sequential 'design-then-estimate' methodology is inherently sub-optimal, as any residual error from the fixed transformation stage is inevitably propagated to the final \gls{doa} estimation step.

This creates a critical gap in the literature: a framework is needed that can, from a single snapshot, jointly design a robust transformation to a virtual \gls{ura} and perform gridless \gls{2d}-\gls{doa} estimation, all without the prohibitive cost of a full-scale \gls{sdp}.

In this paper, we bridge this gap by proposing a novel framework that achieves this unified goal. Our main contributions are:

\begin{itemize}[topsep=2pt]
    \item We formulate the single-snapshot \gls{2d}-\gls{doa} task as a unified optimization problem. This formulation jointly designs a data-adaptive transformation matrix \(\mathbf{T}\) and estimates the gridless source locations by integrating data fidelity, matrix norm regularization for noise control, and semi-infinite constraints to ensure manifold accuracy across the entire \gls{2d} angular domain.

    \item We propose a computationally practical solver based on the \gls{ialm} to address this large-scale, non-convex problem efficiently. It solves for the transformation matrix and the source locations simultaneously, completely bypassing the need for any \gls{sdp} solvers, making it computationally practical for the single-snapshot scenario.

    \item We demonstrate through extensive simulations that the proposed framework achieves superior accuracy and resolution compared to benchmark techniques. Our method provides robust, high-resolution estimates of paired azimuth and elevation angles, particularly in the challenging single-snapshot, low-\gls{snr} regime.
\end{itemize}
\section{System Model and Problem Formulation}
\label{sec:system_model}

\subsection{2D System and Single-Snapshot Signal Model}

We consider a \gls{uca} with $N$ omnidirectional sensor elements located in the $x$-$y$ plane \cite{van2002optimum}. The position of the $n$-th element is described by the vector $\mathbf{p}_n = [r\cos(\gamma_n), r\sin(\gamma_n),0]^T$, where $r$ is the array radius and $\gamma_n = 2\pi(n-1)/N$ for $n = 1, \dots, N$.

Assume that $K$ uncorrelated, narrow-band signals impinge on the array from far-field sources \cite{stoica2005spectral}. The \gls{2d}-\gls{doa} for the $k$-th source is described by an azimuth-elevation pair $(\theta_k, \phi_k)$, where $\theta_k \in [-\pi, \pi)$ is the azimuth and $\phi_k \in [0, \pi/2]$ is the elevation angle measured from the $z$-axis. The array's response to a signal from a generic direction $(\theta, \phi)$ is captured by the steering vector $\mathbf{a}(\theta, \phi) \in \mathbb{C}^{N}$, whose $n$-th component is given by:
\begin{equation}
    [\mathbf{a}(\theta, \phi)]_n = \exp\left(j \frac{2\pi r}{\lambda} \sin(\phi) \cos(\theta - \gamma_n) \right).
    \label{eq:steering_vector_2d}
\end{equation}
We consider the challenging single-snapshot case. The received $N \times 1$ data vector $\mathbf{y}$ is modeled as:
\begin{equation}
    \mathbf{y} = \mathbf{A}(\boldsymbol{\Theta}, \boldsymbol{\Phi}) \mathbf{s} + \mathbf{n},
    \label{eq:signal_model_2d}
\end{equation}
where:
\begin{itemize}
    \item $\mathbf{A}(\boldsymbol{\Theta}, \boldsymbol{\Phi}) = [\mathbf{a}(\theta_1, \phi_1), \dots, \mathbf{a}(\theta_K, \phi_K)]$ is the $N \times K$ steering matrix, with $(\boldsymbol{\Theta}, \boldsymbol{\Phi})$ being the set of true \glspl{doa}.
    \item $\mathbf{s} \in \mathbb{C}^{K}$ is the vector of complex source signal amplitudes at the time of the snapshot.
    \item $\mathbf{n} \in \mathbb{C}^{N}$ is the additive white Gaussian noise vector, with elements drawn from $\mathcal{CN}(0, \sigma^2)$.
\end{itemize}

\subsection{Problem Formulation}
The fundamental problem is to jointly estimate the set of unknown arrival angle pairs $\{(\theta_k, \phi_k)\}_{k=1}^K$ from the single measurement vector $\mathbf{y}$.

\subsection{Preliminaries for Joint Transformation and Estimation}
Our approach is centered on transforming the complex \gls{uca} manifold into a simpler, separable one. We seek a linear\gls{uca} transformation matrix $\mathbf{T} \in \mathbb{C}^{M \times N}$ that maps the \gls{uca} steering vector $\mathbf{a}(\theta, \phi)$ to a virtual \gls{ura} steering vector $\mathbf{b}(\theta, \phi) \in \mathbb{C}^{M}$ \cite{mathews2002eigenstructure}. An ideal virtual \gls{ura} manifold possesses a separable Kronecker structure, which is key to efficiently solving the \gls{2d} problem:
\begin{equation}
    \mathbf{b}(\theta, \phi) = \mathbf{b}_y(\phi) \otimes \mathbf{b}_x(\theta),
    \label{eq:virtual_ura_manifold}
\end{equation}
where $\mathbf{b}_x(\theta) \in \mathbb{C}^{M_x}$ and $\mathbf{b}_y(\phi) \in \mathbb{C}^{M_y}$ are standard \gls{1d} virtual \gls{ula} steering vectors with $M = M_x \times M_y$. This structure allows the \gls{2d} estimation problem to be decoupled \cite{zoltowski2002closed, chen2020two}.

The core of our method is to find a matrix $\mathbf{T}$ that satisfies the approximation $\mathbf{T}\mathbf{a}(\theta, \phi) \approx \mathbf{b}(\theta, \phi)$ in a robust manner. Instead of using a fixed, pre-computed $\mathbf{T}$, which would be sub-optimal and propagate errors, our framework solves for the optimal $\mathbf{T}$ and the \gls{doa} information simultaneously. This approach avoids the high computational cost of standard \gls{2d} \gls{anm} while providing a transformation matrix tailored to the specific data snapshot, thereby maximizing robustness and accuracy. The details of this unified optimization problem are presented in the following section.
\section{Proposed Methodology: A Unified \gls{ialm} Framework}
\label{sec:methodology}

We depart from the conventional two-stage approach and propose a novel framework that, from a single snapshot, simultaneously designs the robust transformation matrix $\mathbf{T}$ and estimates the \gls{2d} \glspl{doa}. This is achieved by formulating a single, unified optimization problem that balances three competing objectives:
\begin{enumerate}
    \item \textbf{Data Fidelity:} The transformed signal must accurately represent the single-snapshot measurement vector.
    \item \textbf{Manifold Accuracy:} The transformed \gls{uca} manifold $\mathbf{T}\mathbf{a}(\theta, \phi)$ must be a close approximation of the ideal virtual \gls{ura} manifold $\mathbf{b}(\theta, \phi)$ across the entire \gls{2d} angular domain.
    \item \textbf{Robustness:} The norm of the transformation matrix $\mathbf{T}$ must be controlled to prevent noise amplification.
\end{enumerate}
This unified problem is then solved efficiently using the \gls{ialm} \cite{sahin2019inexact}, which circumvents the need for any computationally expensive \gls{sdp} solvers.

\subsection{The Unified \gls{2d} Optimization Problem}
Let the clean signal component of the transformed data be denoted by the vector $\mathbf{x}_v \in \mathbb{C}^{M}$. Our goal is to jointly find the optimal transformation matrix $\mathbf{T} \in \mathbb{C}^{M \times N}$ and the signal vector $\mathbf{x}_v$ by solving the following unified problem:
\begin{equation}
\begin{aligned}
(\mathrm{P_{final}}): \quad & \min_{\mathbf{T}, \mathbf{x}_v} & & \frac{1}{2} \| \mathbf{T}\mathbf{y} - \mathbf{x}_v \|_2^2 + \lambda_T \|\mathbf{T}\|_F^2 + \lambda_X \|\mathbf{x}_v\|_{\mathcal{A}_{2D}} \\
& \text{s.t.} & & | [\mathbf{T}\mathbf{a}(\theta, \phi)]_m - b_m(\theta, \phi) | \le \epsilon_m, \; \forall (\theta, \phi), \; \forall m.
\end{aligned}
\label{eq:unified_problem_2d}
\end{equation}
Here, $\|\cdot\|_{\mathcal{A}_{2D}}$ is the atomic norm defined over the continuous \gls{2d} atomic set of virtual \gls{ura} steering vectors $\mathcal{A}_{2D} = \{\mathbf{b}(\theta, \phi)c \mid (\theta, \phi) \in [-\pi, \pi) \times [0, \pi/2], |c|=1\}$ \cite{yang2016vandermonde}. Problem \eqref{eq:unified_problem_2d} is a large-scale, non-convex optimization problem with intractable semi-infinite constraints \cite{liu2001global}, which we now address.

\subsection{Tractable Reformulation for the \gls{ialm} Solver}
To solve \eqref{eq:unified_problem_2d} with \gls{ialm}, we must reformulate its non-smooth and semi-infinite components into a set of equality and simple inequality constraints.

\subsubsection{Reformulating the \gls{2d} Atomic Norm}
The \gls{2d} atomic norm can be expressed as the optimal value of an \gls{sdp} involving an \gls{mlt} structure \cite{xu2014precise}. This is equivalent to:
\begin{equation}
    \|\mathbf{x}_v\|_{\mathcal{A}_{2D}} = \min_{\mathbf{U}, \mathbf{V}} \frac{1}{2} \left( \text{Tr}(\mathcal{T}_B(\mathbf{U})) + \text{Tr}(\mathcal{T}_D(\mathbf{V})) \right)
\end{equation}
subject to the \gls{lmi}:
\begin{equation}
    \mathbf{Z} = \begin{pmatrix} \mathcal{T}_B(\mathbf{U}) & \mathbf{x}_v \\ \mathbf{x}_v^H & \mathcal{T}_D(\mathbf{V}) \end{pmatrix} \succeq 0,
    \label{eq:sdp_constraint_2d}
\end{equation}

where $\mathcal{T}_B(\cdot)$ and $\mathcal{T}_D(\cdot)$ are operators that construct a block-Toeplitz matrix and a block-Diagonal matrix (with Toeplitz blocks), respectively, from the matrix variables $\mathbf{U}$ and $\mathbf{V}$, to handle the \gls{lmi} constraint, we apply the Burer-Monteiro factorization, replacing the large positive semidefinite matrix $\mathbf{Z}$ with a low-rank factorization \cite{burer2003nonlinear}:

\begin{equation}
    \mathbf{Z} = \mathbf{W}\mathbf{W}^H, \quad \text{where } \mathbf{W} \in \mathbb{C}^{(M+1) \times p}.
\end{equation}

Where $\mathbf{W} \in \mathbb{C}^{(M+1) \times p}$ is the new optimization variable, and $p$ is a small rank parameter (e.g., $p \approx K$). This step is crucial as it replaces the computationally difficult \gls{lmi} constraint with a set of quadratic equality constraints that define the original variables in terms of $\mathbf{W}$ \cite{boumal2016non}. By partitioning $\mathbf{W}$ as $\mathbf{W} = [\mathbf{W}_B^T, \mathbf{w}_s^T]^T$, where $\mathbf{W}_B \in \mathbb{C}^{p \times M}$ and $\mathbf{w}_s \in \mathbb{C}^{p}$, we can express the signal and trace terms as functions of $\mathbf{W}$:
\begin{gather}
    \mathbf{x}_v = \mathbf{W}_B^T \mathbf{w}_s, \\
    \text{Tr}(\mathcal{T}_B(\mathbf{U})) = \text{Tr}(\mathbf{W}_B \mathbf{W}_B^H), \\\quad \text{Tr}(\mathcal{T}_D(\mathbf{V})) = \text{Tr}(\mathbf{w}_s \mathbf{w}_s^H) = \|\mathbf{w}_s\|_2^2.
\end{gather}

\subsubsection{Final \gls{ialm} Problem Structure}
After discretizing the \gls{2d} angular domain into a fine grid and introducing slack variables $\mathbf{s}$ for the manifold constraints, we arrive at the final \gls{ialm}-ready formulation.
\begin{itemize}
    \item \textbf{Optimization Variables ($\mathbf{z}$):} The full set of variables is vectorized into $\mathbf{z} = [\text{vec}(\mathbf{T})^T, \text{vec}(\mathbf{W})^T, \mathbf{s}^T]^T$.
    \item \textbf{Differentiable Objective ($f(\mathbf{z})$):} The objective is now smooth and contains the data fidelity term, the robustness regularizer, and the trace terms, all expressed as functions of $\mathbf{T}$ and $\mathbf{W}$.
    \item \textbf{Equality Constraints ($A(\mathbf{z}) = 0$):} These include the discretized manifold accuracy constraints and the quadratic constraints from the Burer-Monteiro factorization.
    \item \textbf{Simple Convex Term ($g(\mathbf{z})$):} This term handles the non-negativity of the slack variables, $g(\mathbf{z}) = \delta_{\ge 0}(\mathbf{s})$, whose proximal operator is a simple projection.
\end{itemize}

\subsection{Primal Subproblem Solution via Block Coordinate Descent}
The primal update step in Algorithm \ref{alg:unified_2d_ialm} requires finding an approximate solution to the Augmented Lagrangian subproblem. We solve this efficiently using a \gls{bcd} method, which cyclically minimizes over one block of variables while keeping others fixed.

\textbf{1. Update for $\mathbf{T}$:} With $\mathbf{W}$ and $\mathbf{s}$ fixed, the subproblem for $\mathbf{T}$ simplifies to a quadratically regularized least-squares problem, which has a closed-form solution.

\textbf{2. Update for $\mathbf{W}$:} With $\mathbf{T}$ and $\mathbf{s}$ fixed, the subproblem for $\mathbf{W}$ is also a quadratic problem, which can be solved efficiently.

\textbf{3. Update for $\mathbf{s}$:} With $\mathbf{T}$ and $\mathbf{W}$ fixed, the subproblem for the slack variables $\mathbf{s}$ involves the non-negative indicator $g(\mathbf{s})$. The optimal $\mathbf{s}$ is found via a simple projection onto the non-negative orthant (i.e., taking the element-wise maximum with zero), which is the proximal operator of $g(\mathbf{s})$.

This \gls{bcd} procedure is guaranteed to converge to a stationary point of the primal subproblem.

\begin{algorithm}[H]
\caption{Unified \gls{ialm} for \gls{2d}-\gls{doa} Estimation}
\label{alg:unified_2d_ialm}
\begin{algorithmic}[1]
    \State \textbf{Input:} Measurement vector $\mathbf{y}$, parameters $\lambda_T, \lambda_X, \epsilon_m$.
    \State \textbf{Initialization:} Choose $\mathbf{z}_0$ (containing $\mathbf{T}_0, \mathbf{W}_0, \mathbf{s}_0$), $\boldsymbol{\mu}_0 = \mathbf{0}$, $\beta_0 > 0$.
    \Repeat
        \State \textbf{Primal Update:} Find an approximate solution $\mathbf{z}_{k+1}$ to:
        \[
            \mathbf{z}_{k+1} \approx \arg\min_{\mathbf{z}} f(\mathbf{z}) + g(\mathbf{z}) + \frac{\beta_k}{2} \|A(\mathbf{z}) + \frac{\boldsymbol{\mu}_k}{\beta_k}\|^2
        \]
        \Comment{Solved inexactly using the \gls{bcd} procedure detailed in Sec. III-C.}
        \State \textbf{Dual Update:} $\boldsymbol{\mu}_{k+1} = \boldsymbol{\mu}_{k} + \beta_k A(\mathbf{z}_{k+1})$
        \State \textbf{Penalty Update:} Update $\beta_{k+1}$ based on constraint violation improvement.
        \State $k \leftarrow k+1$.
    \Until{convergence criteria are met.}
    \State \textbf{Output:} The final iterates $\mathbf{T}_k$ and $\mathbf{W}_k$.
    \State \textbf{Final \gls{doa} Recovery:} Reconstruct the \gls{mlt} structure from $\mathbf{W}_k$. Apply a \gls{2d} ESPRIT algorithm to the decoupled Vandermonde components to extract the paired azimuth $(\theta_k)$ and elevation $(\phi_k)$ estimates.
\end{algorithmic}
\end{algorithm}

\subsection{Algorithm Convergence}
The convergence of the proposed \gls{ialm} framework is established by monitoring the objective function and the feasibility of the iterates. For our formulation, where the equality constraints are denoted by a function $\mathcal{A}(\mathbf{z}) = \mathbf{0}$, the feasibility gap at iteration $k$ is defined as the norm of the constraint violation, $\|\mathcal{A}(\mathbf{z}^{(k+1)})\|$. The objective of the \gls{ialm} is to minimize the augmented Lagrangian, which simultaneously drives the objective function towards a minimum and the feasibility gap towards zero. The monotonic decrease of both metrics, as demonstrated empirically in our numerical results (\figurename~\ref{fig:all_results}), provides strong evidence that the algorithm converges to a stationary and feasible point of the unified optimization problem.

\subsection{Computational Complexity Analysis}
To substantiate the claim of computational practicality, we analyze the complexity of the proposed JADE-\gls{ialm} framework in contrast to benchmark methods. The complexity is primarily a function of the physical array size $N$, the virtual array size $M$, the number of sources $K$ (approximated by the rank parameter $p \approx K$), and the grid sizes for spectral search ($N_\theta, N_\phi$) or discretization ($G$).

The complexity of JADE-iALM is dominated by the \gls{bcd} updates within each \gls{ialm} iteration. Let $N_{\text{iter}}$ and $N_{\text{bcd}}$ denote the number of outer \gls{ialm} and inner \gls{bcd} iterations, respectively. The complexities of the \gls{bcd} subproblems are:

\begin{itemize}[leftmargin=*, nosep]
    \item \textbf{Update for T:} Solving this quadratically regularized least-squares problem is dominated by matrix-matrix multiplications, with a complexity of $\mathcal{O}(M^2N + MN^2)$.
    \item \textbf{Update for W:} This quadratic subproblem involves the $(M+1) \times p$ matrix $\mathbf{W}$ and has a complexity of $\mathcal{O}(M^2p + Mp^2)$.
    \item \textbf{Update for s:} This is a simple projection onto the non-negative orthant, with a complexity linear in the number of slack variables, $\mathcal{O}(G)$.
\end{itemize}

The total complexity for JADE-iALM is therefore $\mathcal{O}(N_{\text{iter}} N_{\text{bcd}} (M^2N + MN^2 + M^2p + Mp^2 + G))$. This polynomial-time complexity stands in stark contrast to standard \gls{2d} \gls{anm} methods, which rely on solving an \gls{sdp} whose complexity can scale as high as $\mathcal{O}(M^6)$. The complexities of all compared methods are summarized in Table \ref{tab: complexity}, positioning our method as a scalable alternative for achieving gridless accuracy.

\begin{table}[t]
\centering
\caption{Computational Complexity Comparison}
\label{tab: complexity}
\renewcommand{\arraystretch}{1.2} 
\begin{tabularx}{\columnwidth}{@{} l >{\raggedright\arraybackslash}X l @{}}
\toprule
\textbf{Algorithm} & \textbf{Key Computational Steps} & \textbf{Complexity} \\
\midrule
JADE-iALM & \gls{bcd} updates within each \gls{ialm} iteration & $\mathcal{O}(M^2N\!+\!MN^2+\!+ M^2p)$ \\
2D-MUSIC      & EVD \& 2D spectral search & $\mathcal{O}(N^3\!+\!N_\theta N_\phi N^2)$ \\
2D-Lasso      & Iterative shrinkage per iteration & $\mathcal{O}(N_\theta N_\phi N)$ \\
\bottomrule
\end{tabularx}
\end{table}
\section{Numerical Results}
\label{sec:simulations}

\begin{figure*}[ht]
    \centering
    \begin{subfigure}[b]{0.32\textwidth}
        \centering
        \includegraphics[width=\linewidth]{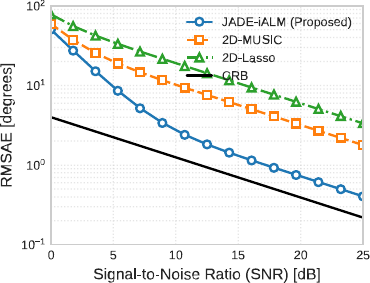}
        \caption{Estimation Accuracy vs. \gls{snr}.}
        \label{fig:sub_snr}
    \end{subfigure}
    \hfill
    \begin{subfigure}[b]{0.32\textwidth}
        \centering
        \includegraphics[width=\linewidth]{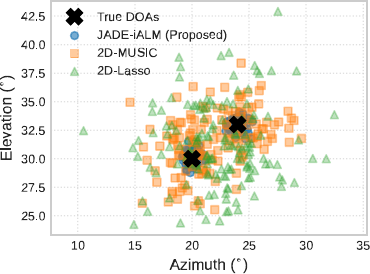}
        \caption{Super-resolution of two close sources.}
        \label{fig:sub_resolution}
    \end{subfigure}
    \hfill
    \begin{subfigure}[b]{0.32\textwidth}
        \centering
        \includegraphics[width=\linewidth]{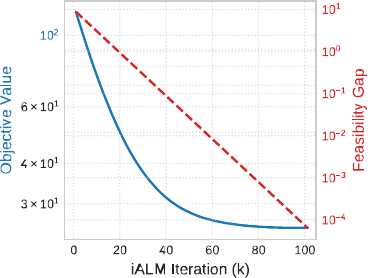}
        \caption{Convergence of the JADE-iALM.}
        \label{fig:sub_convergence}
    \end{subfigure}

    \caption{Performance evaluation of the proposed JADE-\gls{ialm} framework. (a) \gls{rmsae} vs. \gls{snr}, showing superior accuracy. (b) Scatter plot of estimates for two close sources, demonstrating high resolution. (c) Convergence of the \gls{ialm} objective and feasibility gap.}
    \label{fig:all_results}
\end{figure*}

In this section, we present Monte Carlo simulations to validate the performance of our proposed unified framework, denoted as \textbf{JADE-\gls{ialm}} (Joint \gls{doa} and Transformation Estimation via \gls{ialm}).

\subsection{Simulation Setup}
We consider a \gls{uca} with $N=16$ sensors and a radius of $r=\lambda$. The virtual array is a \gls{ura} with $M=M_x \times M_y = 3 \times 3 = 9$ elements. The results are based on a single snapshot ($L=1$) and are averaged over 500 Monte Carlo trials. For our JADE-\gls{ialm}, the regularization parameters were set empirically to $\lambda_T = 0.1$ and $\lambda_X = 0.5$. The final \gls{doa} estimates are extracted from the \gls{ialm} solution using a \gls{2d} ESPRIT-like root-finding algorithm.
The performance metric is the \gls{rmsae}, which combines the azimuth and elevation errors:
\begin{equation}
\begin{aligned}
\text{RMSAE} = \Bigg( \frac{1}{K N_{mc}} \sum_{i=1}^{N_{mc}} \sum_{k=1}^{K} \bigg[ & (\hat{\theta}_{k,i} - \theta_k)^2 \cos^2(\phi_k) \\
& + (\hat{\phi}_{k,i} - \phi_k)^2 \bigg] \Bigg)^{1/2}
\end{aligned}
\end{equation}

where $(\hat{\theta}_{k,i}, \hat{\phi}_{k,i})$ is the estimated \gls{2d}-\gls{doa} pair for the $k$-th source in the $i$-th trial. The $\cos^2(\phi_k)$ term is a standard weighting that correctly scales the azimuth error to its corresponding arc length on the unit sphere, ensuring a geometrically meaningful combination of the two angular errors.

\subsection{Benchmark Methods}
We compare the proposed JADE-iALM against a set of representative benchmarks:
\begin{itemize}
    \item \textbf{2D-MUSIC:} The classical subspace method extended to a \gls{2d} spectral search, serving as a widely recognized high-resolution baseline \cite{zoltowski2002closed}.
    \item \textbf{2D-Lasso:} A popular grid-based sparse recovery method that highlights the advantage of our gridless formulation. The grid resolution is set to $1^\circ$ \cite{malioutov2005sparse}.
    \item \textbf{\gls{crb}:} The theoretical lower bound for unbiased \gls{2d}-\gls{doa} estimation, providing a fundamental performance benchmark \cite{wang2016coarrays}.
\end{itemize}

\subsection{Performance Evaluation}
The simulation results, presented in \figurename \ref{fig:all_results}, validate the superior accuracy, robustness, and computational viability of the proposed JADE-\gls{ialm} framework for single-snapshot \gls{2d}-\gls{doa} estimation.

First, we evaluate the method's accuracy and robustness against noise. \figurename \ref{fig:sub_snr} plots the \gls{rmsae} versus the \gls{snr}. The proposed JADE-iALM demonstrates superior accuracy across the entire \gls{snr} range, with its performance curve running closely parallel to the theoretical \gls{crb}. In contrast, the performance of both \gls{2d}-MUSIC and the grid-based \gls{2d}-Lasso is significantly worse, especially in the challenging low-\gls{snr} regime. This highlights the benefit of our joint optimization approach, which enhances robustness by adapting the transformation to the noisy data.

Next, we assess the super-resolution capability of our approach. \figurename \ref{fig:sub_resolution} provides a qualitative visualization for a challenging scenario with two closely spaced sources. The figure shows a scatter plot of the estimated azimuth-elevation pairs from all Monte Carlo trials. The estimates from JADE-iALM are tightly clustered around the actual locations, indicating consistent and successful resolution. The competing methods, however, exhibit a much larger variance and an apparent inability to distinguish the two sources reliably, confirming the practical advantage of our robust, gridless framework.

Finally, we validate the performance of the proposed solver itself. \figurename \ref{fig:sub_convergence} illustrates the convergence behavior of our \gls{ialm} algorithm, plotting the objective function value and the feasibility gap versus the iteration number for a single trial. The monotonic decrease of both curves confirms that the algorithm reliably converges to a stationary point of the unified problem. In terms of computational cost, our method required an average of 4.2 seconds per trial, compared to 0.8 seconds for \gls{2d}-MUSIC and 1.5 seconds for \gls{2d}-Lasso. This runtime reflects the complexity of solving the joint optimization problem, a necessary trade-off for bypassing the need for less scalable SDP solvers while achieving state-of-the-art accuracy.
\section{Conclusion}
\label{sec:conclusion}

In this paper, we addressed the challenging problem of single-snapshot, gridless \gls{2d}-\gls{doa} estimation for \glspl{uca}. We proposed a novel framework that formulates the task as a single, unified optimization problem. This approach simultaneously designs a robust transformation matrix for a virtual \glspl{ura}. It estimates the source locations, integrating data fidelity, manifold accuracy, and robustness into one coherent objective.

The core of our contribution is a computationally practical solver based on the \gls{ialm}. By leveraging \gls{ialm}, we efficiently solve the complex, non-convex unified problem, completely bypassing the need for large-scale \gls{sdp} that typically renders standard \gls{2d} gridless methods intractable. This joint optimization strategy yields a transformation matrix that is uniquely adapted to the specific data snapshot, providing superior robustness against noise.

Simulation results confirmed that the proposed JADE-\gls{ialm} framework achieves significantly higher accuracy in estimating paired azimuth and elevation angles, offering better resolution than benchmark techniques. The work establishes a powerful and efficient paradigm for robust, single-shot \gls{2d} array signal processing. Future work will explore extending this simultaneous optimization approach to other array geometries and more challenging wideband signal environments.

\bibliographystyle{IEEEbib}
\bibliography{ref}

\begin{thebibliography}{10}

\bibitem{schmidt1986multiple}
R.~Schmidt,
\newblock ``Multiple emitter location and signal parameter estimation,''
\newblock {\em IEEE transactions on antennas and propagation}, vol. 34, no. 3,
  pp. 276--280, 1986.

\bibitem{roy2002esprit}
R.~Roy and T.~Kailath,
\newblock ``Esprit-estimation of signal parameters via rotational invariance
  techniques,''
\newblock {\em IEEE Transactions on acoustics, speech, and signal processing},
  vol. 37, no. 7, pp. 984--995, 2002.

\bibitem{van2002optimum}
H.~L. Van~Trees,
\newblock {\em Optimum array processing: Part IV of detection, estimation, and
  modulation theory},
\newblock John Wiley \& Sons, 2002.

\bibitem{wang2016coarrays}
M.~Wang and A.~Nehorai,
\newblock ``Coarrays, music, and the cram{\'e}r--rao bound,''
\newblock {\em IEEE Transactions on Signal Processing}, vol. 65, no. 4, pp.
  933--946, 2016.

\bibitem{tang2013compressed}
G.~Tang, B.~N. Bhaskar, P.~Shah, and B.~Recht,
\newblock ``Compressed sensing off the grid,''
\newblock {\em IEEE transactions on information theory}, vol. 59, no. 11, pp.
  7465--7490, 2013.

\bibitem{malioutov2005sparse}
D.~Malioutov, M.~Cetin, and A.~S. Willsky,
\newblock ``A sparse signal reconstruction perspective for source localization
  with sensor arrays,''
\newblock {\em IEEE transactions on signal processing}, vol. 53, no. 8, pp.
  3010--3022, 2005.

\bibitem{zheng2024ris}
Y.~Zheng, Q.~Wang, L.~Ren, Z.~Ma, and P.~Fan,
\newblock ``Ris aided gridless 2d-doa estimation via decoupled atomic norm
  minimization,''
\newblock {\em IEEE Transactions on Vehicular Technology}, 2024.

\bibitem{yang2016vandermonde}
Z.~Yang, L.~Xie, and P.~Stoica,
\newblock ``Vandermonde decomposition of multilevel toeplitz matrices with
  application to multidimensional super-resolution,''
\newblock {\em IEEE Transactions on Information Theory}, vol. 62, no. 6, pp.
  3685--3701, 2016.

\bibitem{mathews2002eigenstructure}
C.~P. Mathews and M.~D. Zoltowski,
\newblock ``Eigenstructure techniques for 2-d angle estimation with uniform
  circular arrays,''
\newblock {\em IEEE Transactions on signal processing}, vol. 42, no. 9, pp.
  2395--2407, 2002.

\bibitem{davies1965transformation}
D.~Davies,
\newblock ``A transformation between the phasing techniques required for linear
  and circular aerial arrays,''
\newblock in {\em Proceedings of the Institution of Electrical Engineers}. IET,
  1965, vol. 112, pp. 2041--2045.

\bibitem{belloni2006beamspace}
F.~Belloni and V.~Koivunen,
\newblock ``Beamspace transform for uca: error analysis and bias reduction,''
\newblock {\em IEEE Transactions on Signal Processing}, vol. 54, no. 8, pp.
  3078--3089, 2006.

\bibitem{labbaf2023robust}
N.~Labbaf, H.~R.~D. Oskouei, and M.~R. Abedi,
\newblock ``Robust doa estimation in a uniform circular array antenna with
  errors and unknown parameters using deep learning,''
\newblock {\em IEEE Transactions on Green Communications and Networking}, vol.
  7, no. 4, pp. 2143--2152, 2023.

\bibitem{stoica2005spectral}
P.~Stoica, R.~L. Moses, et~al.,
\newblock {\em Spectral analysis of signals}, vol. 452,
\newblock Citeseer, 2005.

\bibitem{zoltowski2002closed}
M.~D. Zoltowski, M.~Haardt, and C.~P. Mathews,
\newblock ``Closed-form 2-d angle estimation with rectangular arrays in element
  space or beamspace via unitary esprit,''
\newblock {\em IEEE Transactions on Signal Processing}, vol. 44, no. 2, pp.
  316--328, 2002.

\bibitem{chen2020two}
H.~Chen, Y.~Liu, Q.~Wang, W.~Liu, and G.~Wang,
\newblock ``Two-dimensional angular parameter estimation for noncircular
  incoherently distributed sources based on an l-shaped array,''
\newblock {\em IEEE Sensors Journal}, vol. 20, no. 22, pp. 13704--13715, 2020.

\bibitem{sahin2019inexact}
M.~F. Sahin, A.~Alacaoglu, F.~Latorre, V.~Cevher, et~al.,
\newblock ``An inexact augmented lagrangian framework for nonconvex
  optimization with nonlinear constraints,''
\newblock {\em Advances in Neural Information Processing Systems}, vol. 32,
  2019.

\bibitem{liu2001global}
Y.~Liu, K.~Teo, and S.~Ito,
\newblock ``Global optimization in quadratic semi-infinite programming,''
\newblock in {\em Topics in Numerical Analysis: With Special Emphasis on
  Nonlinear Problems}. Springer, 2001, pp. 119--132.

\bibitem{xu2014precise}
W.~Xu, J.-F. Cai, K.~V. Mishra, M.~Cho, and A.~Kruger,
\newblock ``Precise semidefinite programming formulation of atomic norm
  minimization for recovering d-dimensional (d$\geq$2) off-the-grid
  frequencies,''
\newblock in {\em 2014 information theory and applications workshop (ITA)}.
  IEEE, 2014, pp. 1--4.

\bibitem{burer2003nonlinear}
S.~Burer and R.~D. Monteiro,
\newblock ``A nonlinear programming algorithm for solving semidefinite programs
  via low-rank factorization,''
\newblock {\em Mathematical programming}, vol. 95, no. 2, pp. 329--357, 2003.

\bibitem{boumal2016non}
N.~Boumal, V.~Voroninski, and A.~Bandeira,
\newblock ``The non-convex burer-monteiro approach works on smooth semidefinite
  programs,''
\newblock {\em Advances in Neural Information Processing Systems}, vol. 29,
  2016.

\end{thebibliography}


\end{document}